\documentclass[a4paper,fleqn,usenatbib]{mn2e}

\usepackage{times}

\usepackage{graphicx}

\newif\ifAMStwofonts

\def\gtorder{\mathrel{\raise.3ex\hbox{$>$}\mkern-14mu
             \lower0.6ex\hbox{$\sim$}}}
\def\ltorder{\mathrel{\raise.3ex\hbox{$<$}\mkern-14mu
             \lower0.6ex\hbox{$\sim$}}}

\title{Faint, high proper motion star candidates selected from the SDSS and PS1 catalogs}

\author[Segev \& Ofek]
       {Noam Segev$^{1}$,  Eran O. Ofek\thanks{e-mail: eran.ofek@weizmann.ac.il}$^{1}$ \\
$^{1}$ Benoziyo Center for Astrophysics, Weizmann Institute of Science, 76100 Rehovot, Israel}

\date{Accepted ?
      Received ?
      in original form ?}

\begin{document}

\maketitle

\begin{abstract}

High proper motion stars probe several extreme stellar populations, including nearby objects,
Galactic halo stars, and hyper-velocity stars.
Extending the search for high proper motion stars, to faint limits can increase
their numbers and help to identify interesting targets.
We conduct a search for faint ($r_{\rm SDSS}>19.5$\,mag) high proper motion stars
($\gtorder 200$\,mas\,yr$^{-1}$)
by comparing the Sloan Digital Sky Survey (SDSS) - Data Release (DR) 10 catalog to
the Pan-STARRS1-DR1 stacked image catalog.
Our main selection criterion is stars that moved $>1\farcs{5}$ 
and up to $7''$ between the SDSS and PS1 epochs.
We identify 2923 high proper motion stars, of which 826 do~not have proper motion
in the GAIA-DR2 catalog and 565 are not listed in the GAIA-DR2 catalog.
Our SDSS-PS1 proper motions are consistent with the GAIA-measured
proper motions with a robust rms of about 10\,mas\,yr$^{-1}$.

\end{abstract}


\section{Introduction}
\label{sec:intro}

High proper motion stars probe several stellar populations.
They signal out candidates for nearby objects; probe halo stars and
allow us to estimate
the halo age using white dwarfs (WD; e.g., Kilic et al. 2017);
estimate the galactic escape velocity and gravitational potential
(e.g.,
Gnedin et al. 2005; Yu \& Madau 2007, Gnedin et al. 2010; Rossi et al. 2017;
Fragione \& Loeb 2017); and
discover high-velocity stars, which are presumably formed in
three body interactions with the Galactic center black hole (e.g., Brown 2015);
as well as identify inter Galactic interlopers
(e.g., Erkal et al. 2018;
Marchetti, Rossi \& Brown 2018).
In addition, the nearest isolated neutron star may hide
among faint high proper motion stars (e.g., Ofek 2009).
Furthermore, a recent search suggests that some high-velocity stars
have unusual spectra (Shen et al. 2018).

The recent GAIA-DR2 release (Gaia Collaboration et al. 2016b;
Gaia Collaboration et al. 2018) presents proper motion and parallax fits
for about 1.3 billion sources.
The GAIA-DR2 catalog still suffers from a high fraction of missing
high proper motion stars. This is expected to be improved as more epochs
will become available.
Tian et al. (2017)
presented a proper motion catalog of about $3.5\times10^{8}$ stars based on merging
the GAIA-DR1 (Gaia Collaboration et al. 2016a), Pan-STARRS1-DR1 (Chambers et al. 2016), SDSS catalog (York et al. 2000),
and the 2MASS catalog (Cutri, et al. 2003).
Their search is limited to sources that moved up to $1\farcs{5}$
between the SDSS and PS1 epochs (typically $\ltorder0.2$\,mas\,yr$^{-1}$).

Here we present a search for faint ($r$-band mag $>19.5$) high proper motion
objects that moved $>1\farcs{5}$ and $<7''$ between the SDSS and PS1 epochs.
Our search is not designed to be complete, and its main purpose is
to generate a list of interesting targets for follow-up observations.
This source-list may be important for searches of high-velocity stars,
cool objects and cool stellar remnants.

In \S\ref{sec:search} we present the search criteria and methodology.
In \S\ref{sec:cand} we list the candidates and in \S\ref{sec:disc} we discuss the results.

\section{Search for proper motion stars}
\label{sec:search}

We used the {\tt catsHTM} tool (Soumagnac \& Ofek 2018) to query
the SDSS-DR10 (Ahn et al. 2014), PS1-DR1 (Chambers et al. 2016;
Magnier et al. 2016a; 2016b; 2016c), and
GAIA-DR2  (Gaia Collaboration et al. 2016b;
Gaia Collaboration et al. 2018) catalogs.
We selected SDSS sources that are of type=6 (unresolved);
whose angular velocities,
as measured in a single SDSS image,
are smaller than three times the uncertainty in
their velocities (i.e., not an asteroid);
and that have $r_{\rm SDSS}>19.5$\,mag.
For each such source, we searched
for PS1-DR1 (Chambers et al. 2016;
Magnier et al. 2016a; 2016b; 2016c)
sources in its neighborhood.
We selected SDSS sources that do~not have counterparts
in the PS1 stacked-image catalog within $1\farcs{5}$.
The PS1 stacked-image catalog has
the advantage of beeing deeper than single
epoch PS1 images.
However, due to the several years baseline of the PS1
survey, in many cases proper motion stars
are elongated in the stacked image
and their positional uncertainty is larger than
the typical astrometric errors in the catalog.

Next, we searched for PS1 sources around the SDSS source,
and selected cases in which there is exactly one PS1 source within $7''$
from the SDSS source.
We also checked that this PS1 source does~not have an SDSS
counterpart, within $1.5''$.
Although, this may reject some high proper motion
stars (especially at low Galactic latitudes),
this step removes a large number of
errors and cases of confusion that are hard
to verify using only two epochs.
Next, we checked that the
SDSS $X$-band magnitude is brighter than 21.7\,magnitude and that the PS1 $X$-band
magnitude is brighter than 22.1\,mag.
Here, $X$-band is either $g$-, $r$-, or $i$-band and we demanded
that this criterion is fulfilled for at least one band.
We further selected stars for which the best SDSS magnitude error in one
of the five SDSS bands is smaller than 0.12\,mag, and
the second best magnitude error is smaller than 0.2\,mag.
These somewhat arbitrary cuts were very useful in removing large
number of problems including spurious sources
(e.g.,
background fluctuations that looks like stars)
and image artifacts.

Following this, we selected all SDSS and PS1 sources within $200''$
of the candidate position,
we matched them by coordinates and calculated the mean offset
in right ascension (R.A.) and declination (Dec.)
between the two matched lists.
If the absolute value of local offset in R.A. was smaller than $0\farcs{5}$,
and the absolute value of local offset in Dec. was smaller than $0\farcs{5}$,
we proceeded; otherwise, we rejected the candidate.
This step rejected a large number of false candidates
due to bad astrometry.

Next we checked that the $|X_{\rm SDSS} - X_{\rm PS1}|$ magnitude difference,
is smaller than 0.4\,mag in at least one of the bands, where
$X$ is either $g$, $r$, or $i$-bands.
Finally, we removed sources which have another SDSS source
brighter than $r$-band 16\,magnitude within $20''$,
or brighter than $r$-band 13.5\,magnitude within $50''$.
This step rejected large number of fake sources that are found
near diffraction spikes and saturated stars.

This selection process yielded 4486 initial candidates for high proper-motion stars.
We inspected by eye the SDSS and PS1 images of all the candidates.
Sometime, when more than one SDSS image was available,
we inspected multiple SDSS images.
This inspection revealed many false candidates.
The most common problems were due to
faint sources that were not identified in one of the images,
or sources that are likely due to noise,
diffraction spikes and saturated stars,
features on top of extended sources, like galaxies and nebulae,
that were misidentified as stars, bad astrometry,
and transients.
All the steps in this search were conducted using
tools available as part of the MATLAB Astronomy \& Astrophysics
Toolbox\footnote{https://webhome.weizmann.ac.il/home/eofek/matlab/}
(Ofek 2014).
After this selection process, we were left with 2923 good 
candidates for high proper motion stars.

\section{The high proper motion candidates}
\label{sec:cand}

Our search yielded 2923 high proper motion candidates, listed in Table~\ref{tab:list}.
We cross-matched these sources with the GAIA-DR2 catalog.
Of these 2923, 826 stars do~not have proper motion measurements in the GAIA-DR2 catalog,
and 565 are not listed in the GAIA-DR2 catalog.
644 stars in this list have SDSS spectra,
and their spectral type information is listed in Table~\ref{tab:list}.

\begin{table*}
\begin{tabular}{ccccccccccccc}
\hline
\multicolumn{2}{c}{J2000 SDSS} &     &                &              &      &      & \multicolumn{6}{c}{SDSS}\\
$\alpha$ & $\delta$ & $\mu_{\alpha}$ & $\mu_{\delta}$ & Epoch        & Sep. & Time & $g$ & $\delta{g}$ & $r$ & $\delta{r}$ & $i$ & $\delta{i}$ \\
deg      & deg      & mas\,yr$^{-1}$ & mas\,yr$^{-1}$ & Julian years & $''$& yr   & mag & mag         & mag & mag             & mag & mag \\
\hline
358.1971914 & $   -20.1561758$  & $ -199.3$ & $ -109.1$ &  2004.951 &   1.77 &  7.77  & 24.34 &0.68 &22.84 &0.33 &19.77 &0.04 \\
355.0348332 & $   -19.6746000$  & $  144.0$ & $ -119.1$ &  2004.951 &   1.82 &  9.71  & 21.46 &0.06 &20.13 &0.03 &18.32 &0.01 \\
344.2284300 & $   -19.1558072$  & $  298.0$ & $  -19.4$ &  2004.951 &   2.23 &  7.45  & 24.76 &0.70 &22.06 &0.14 &19.60 &0.03 \\
312.6912417 & $   -15.6465449$  & $   63.6$ & $ -315.6$ &  2004.697 &   2.17 &  6.73  & 22.55 &0.17 &20.80 &0.05 &18.19 &0.01 \\
304.1203511 & $   -13.1012241$  & $   -4.7$ & $ -292.8$ &  2005.441 &   1.77 &  6.03  & 23.68 &0.57 &21.88 &0.17 &18.61 &0.02 \\
\hline
\end{tabular}
\caption{{\bf List of candidate faint high proper motion stars}. The full table is available electronically.
Here we give the first five lines, and first 13 columns. Additional columns available in the electronic version
are the PS1 $gri$ magnitudes, a flag indicating if a common proper motion star was detected
in the images, number of GAIA matches, GAIA proper motion and parallax, SDSS spectroscopic classification
and properties.}
\label{tab:list}
\end{table*}

\begin{figure}
\includegraphics[width=80mm]{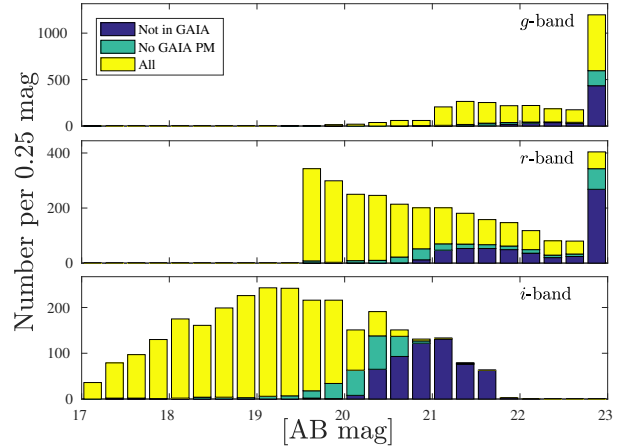}
\caption{Proper motion star candidates' $g$ (top), $r$ (middle), and $i$ (bottom) SDSS AB-magnitude distribution.
The highest-magnitude bin (on the right side) in the $g$- and $r$-bands represents magnitudes fainter than
the SDSS detection limit.}
\label{fig:Cand_Mag_hist}
\end{figure}

\begin{figure}
\includegraphics[width=80mm]{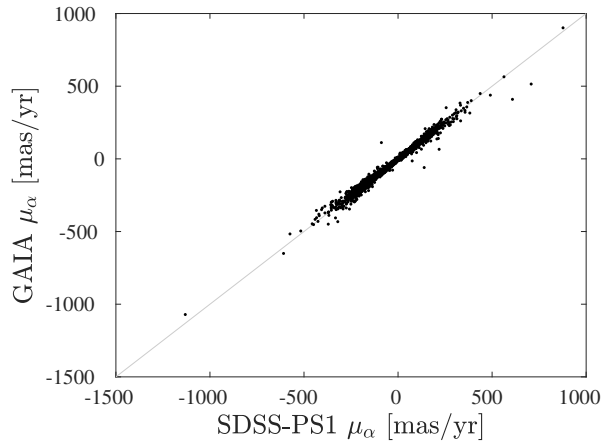}
\caption{SDSS-PS1 vs. GAIA proper motion in Right Ascension ($\alpha$) for our high proper motion
stars that have proper motion measurements in the GAIA catalog.}
\label{fig:PM_RA_sdss_gaia}
\end{figure}

\begin{figure}
\includegraphics[width=80mm]{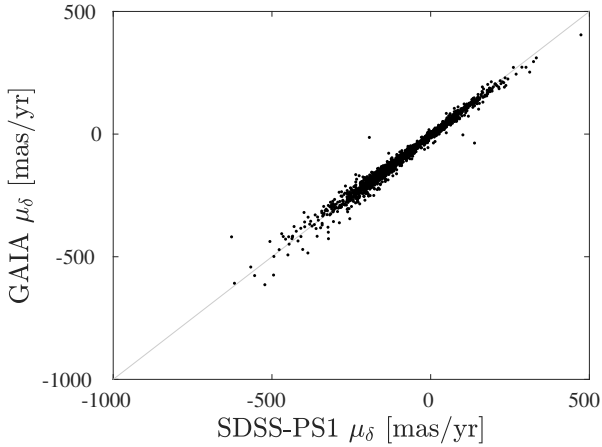}
\caption{Like Figure~\ref{fig:PM_RA_sdss_gaia} but for the proper motion in declination ($\delta$).}
\label{fig:PM_Dec_sdss_gaia}
\end{figure}

Figure~\ref{fig:Cand_Mag_hist} presents the candidates' magnitude distribution.
To verify our results, we show in Figures~\ref{fig:PM_RA_sdss_gaia} and \ref{fig:PM_Dec_sdss_gaia} the SDSS-PS1 proper motion
vs. the GAIA proper motion, in Right Ascension ($\alpha$) and Declination ($\delta$), respectively.
There is good agreement between the proper motion measurements.
Assuming that the GAIA measurement errors are much smaller than our uncertainties
(see justification below),
these figures also allow us to estimate the uncertainty in our measurements.
We find that for
the proper motion in right-ascension (declination),
there is a small bias of about 1 (6)\,mas\,yr$^{-1}$ between
the SDSS-PS1 and GAIA measurements, and that the scatter is 19 (17)\,mas\,yr$^{-1}$.
This scatter is dominated by a few outliers, and the
robust scatter\footnote{We estimate the robust scatter by calculating the $15.9$ to $84.1$ percentiles of the distribution divided by two.}
is about 10\,mas\,yr$^{-1}$.
These estimates are consistent with the expected uncertainty.
Specifically, the typical astrometric uncertanty in
the PS1 and SDSS images is of the order of 50--100\,mas.
Adding in quadrature and dividing by the typical time baseline
between the two surveys gives expected errors of the order of 10\,mas\,yr$^{-1}$.
Note that the errors in astrometry can likely be improved
by redoing the astrometry of the PS1 and SDSS images
using the GAIA catalog (e.g., Tian et al. 2017).

\section{Discussion}
\label{sec:disc}

\begin{figure}
\includegraphics[width=80mm]{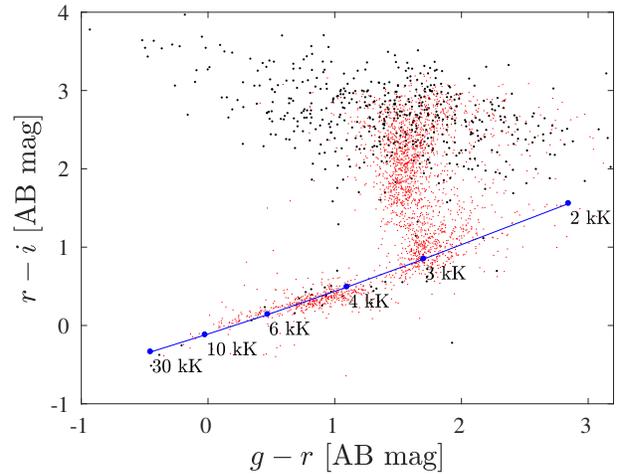}
\caption{SDSS $g-r$ vs. $r-i$ colors of the high proper motion candidates.
Tiny red points represent sources with reduced proper motion of $H<24$\,mag,
while black points have $H\ge24$\.mag.
The blue line represent the expected color of a black body source, assuming no reddening.
There are many sources (mainly black points) that are not located on the black-line.
These are sources whose magnitude in one of the bands is below the detection limit
and, therefore, one of the colors is unreliable.}
\label{fig:gr_ri}
\end{figure}
\begin{figure}
\includegraphics[width=80mm]{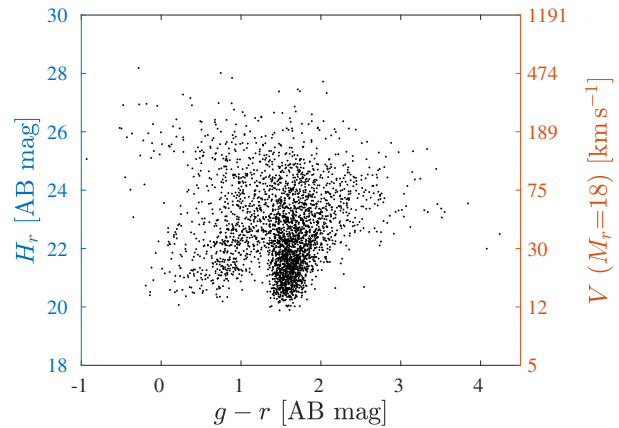}
\caption{The $g-r$ color vs. reduced proper motion of high proper motion candidates.
The right-hand axis shows the transverse velocity assuming absolute $r$-band magnitude of 18.}
\label{fig:gr}
\end{figure}
\begin{figure}
\includegraphics[width=80mm]{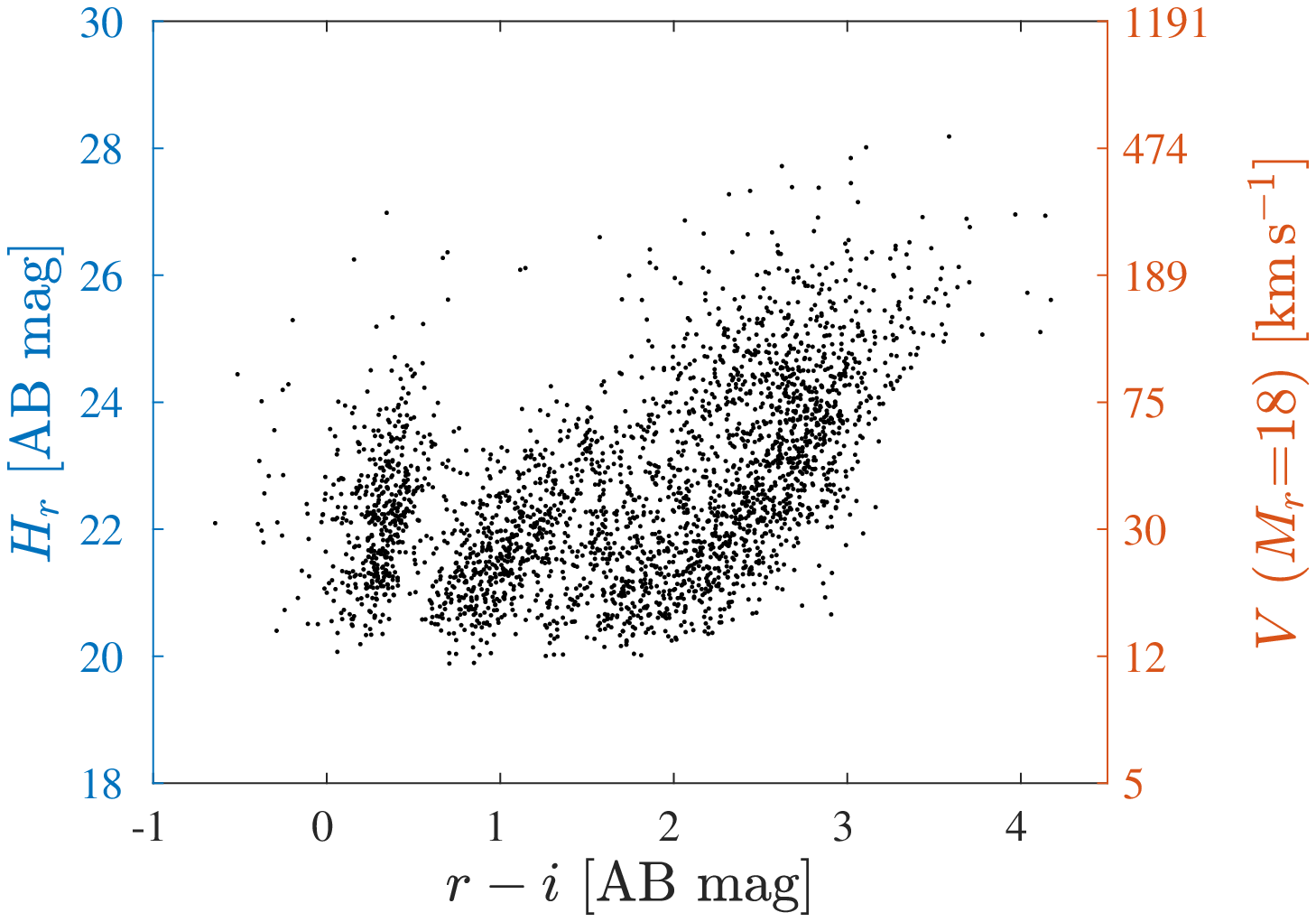}
\caption{Like Figure~\ref{fig:gr}, but for the $r-i$ color.}
\label{fig:ri}
\end{figure}

We present a non-complete search for faint high proper motion stars from the SDSS and PS1 catalogs.
Our search yielded 2923 high proper motion candidates listed in Table~\ref{tab:list},
of which 826 stars do~not have proper motion measurements in the GAIA-DR2 catalog.
This is the largest list of high proper motion stars below the USNO-B1.0 (Monet et al. 2003)
and GAIA magnitude limits.

Figure~\ref{fig:gr_ri} shows the SDSS $g-r$ vs. $r-i$ colors of these sources.
Red points represent sources with a reduced proper motion of a $H<24$\,mag,
while black points have $H\ge24$\.mag.
Here, the reduced proper motion is defined as
\begin{equation}
H = m + 5\log_{10}(\mu) + 5 = M + 5\log_{10}{V} - 3.379,
\label{eq:H}
\end{equation}
where $M$ is the absolute magnitude, $\mu$ is the total proper motion in arcseconds per year,
and $V$ is the transverse velocity in km\,s$^{-1}$.
The blue line represents the expected color of a black body source, assuming no reddening.
In this plot, hot stars, with an effective temperature above 4000\,K, are presumably white dwarfs,
while the vertical clump is likely populated by M dwarfs and brown dwarfs.
There are many sources (mainly black points; i.e., $H\ge24$) that are not located on the black-line.
These are sources whose magnitude in one of the bands is below the detection limit
and, therefore, one of the colors is unreliable.

\begin{figure}
\includegraphics[width=80mm]{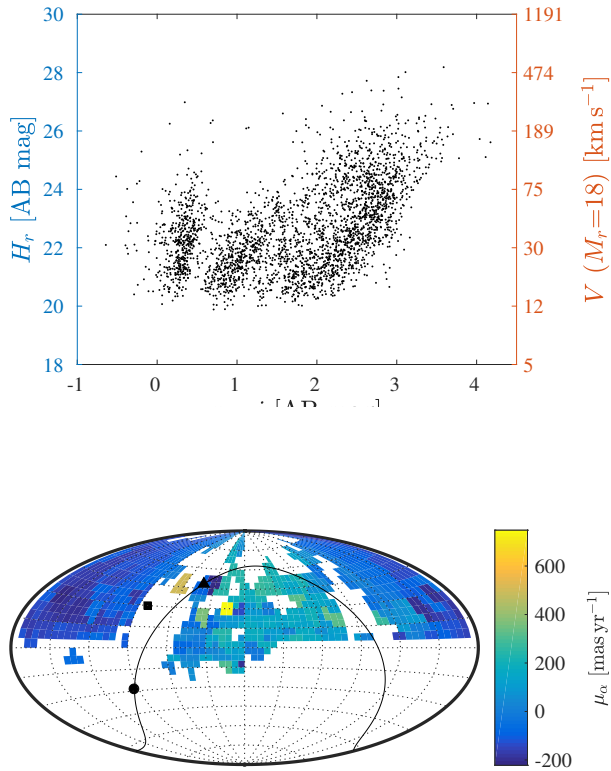}
\caption{Sky map in Aitoff projection and Equatorial coordinates,
of the mean proper motion in Right Ascension in our catalog of 2923 high proper motion stars.
The mean proper motion was calculated in 512 hierarchical triangular
mesh trixels
and then interpolated, using nearest point interpolation, to a regular grid.
The solid line represents the Galactic plane,
The black circle, triangle, and square, shows the Galactic center,
the direction of motion of the solar circle, 
and the direction of the Solar Apex ($\alpha=277$\,deg, $\delta=+30$\,deg),
respectively.\vspace{8mm}}
\label{fig:Map_PM_RA}
\end{figure}

\begin{figure}
\includegraphics[width=80mm]{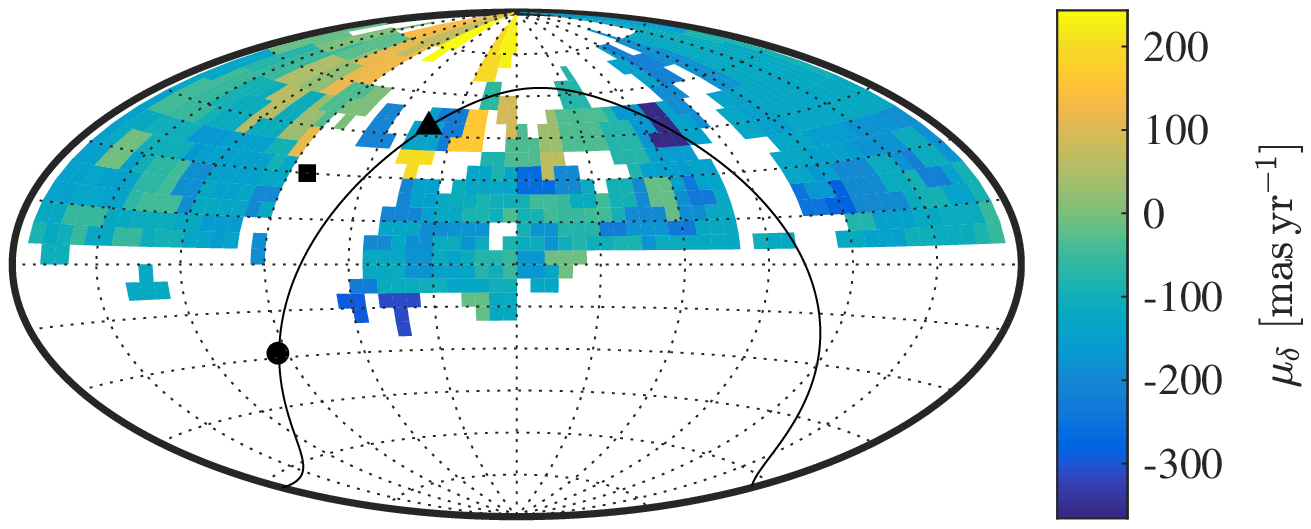}
\caption{Like Figure~\ref{fig:Map_PM_RA}, but for proper motion in declination.}
\label{fig:Map_PM_Dec}
\end{figure}

\begin{figure}
\begin{center}
\includegraphics[width=70mm]{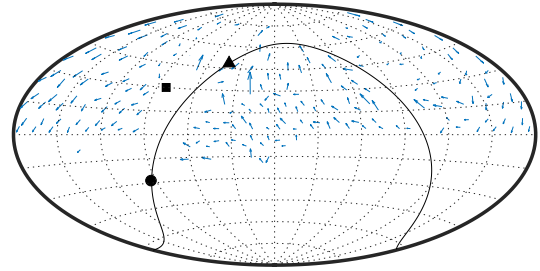}
\end{center}
\caption{Like Figure~\ref{fig:Map_PM_RA}, but a quiver map of proper motions,
where arrows indicate mean proper motion direction and amplitude, for each trixel
in the hierarchical triangular mesh.}
\label{fig:Map_PM_quiver}
\end{figure}

Figure~\ref{fig:gr} (\ref{fig:ri}) presents the SDSS $g-r$ ($r-i$) color of these stars against their reduced proper motion.
The right-hand Y-axis shows the transverse velocity assuming an absolute magnitude of 18.
Figure~\ref{fig:Map_PM_RA} (\ref{fig:Map_PM_Dec}) shows the mean proper motion in right ascension (declination)
as a function of sky location as calculated in 512 hierarchical triangular mesh
(Szalay et al. 2007) trixels
(triangle side is about 11\,deg).
These plots suggest that the proper motion component of faint high proper motion
stars are not random and show some structure.
Figure~\ref{fig:Map_PM_quiver} presents
a quiver map of proper motion,
where arrows indicate mean proper motion direction and amplitude, for each trixel in the hierarchical triangular mesh.


\begin{figure}
\begin{center}
\includegraphics[width=70mm]{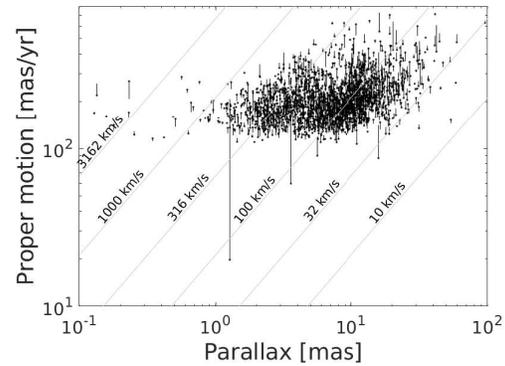}
\end{center}
\caption{The parallax vs.
proper motion of stars in our sample that have GAIA-DR2 proper motions.
The difference between a GAIA proper motion
measurement and our SDSS/PS1-based measurement is indicated by the vertical lines.
The grey curves show lines of equal sky-projected velocities.}
\label{fig:GAIA_plx_pm}
\end{figure}

An interesting question is what fraction of these new high proper motion stars are
high velocity stars, or nearby objects.
This question cannot be answered directly. However,
we can get an estimated answer by inspecting the $\approx2000$ stars in
our sample for which GAIA parallaxes are available.
Figure~\ref{fig:GAIA_plx_pm} presents, for each star in this subset, the parallax vs.
the measured proper motion. The differences between the GAIA proper motion
measurements and our SDSS/PS1-based measurements are indicated by the vertical lines.
The grey curves show lines of equal sky-projected velocities.
Presumably most of these faint high proper motion stars are relatively nearby ($\ltorder1$\,kpc) stars.
However, about 30 stars (11 with negative parallaxes and hence are not shown in the plot),
have nominal projected velocities consistent with being above 1000\,km\,s$^{-1}$.
However, the median error in GAIA parallax is about 0.5\,mas, and we cannot rule out
that most of the these high velocities are due to errors.
Therefore, we estimate that up to about 2\% of the stars in our sample may have
projected velocities above about 1000\,km\,s$^{-1}$.

To summarize, our new list of faint proper motion stars
contains candidates for hyper velocity stars,
halo stars, cool stars, and nearby objects.
This list provide an order of magnitude increase
in the number of known high proper motion stars fainter
than $r$-band magnitude $\approx20.5$.

\section*{ACKNOWLEDGMENTS}
E.O.O. is grateful for the support by
grants from the Israel Science Foundation, Minerva, Israeli Ministry of Technology and Science, the US-Israel Binational Science Foundation, Weizmann-UK,
and the I-CORE Program of the Planning and Budgeting Committee and the Israel Science Foundation.



\end{document}